\documentclass[conference]{IEEEtran}

\usepackage{cite}
\usepackage[utf8]{inputenc}
\usepackage{amsmath,amssymb,amsfonts}
\usepackage{textcomp}

\usepackage{graphicx}
\usepackage{xcolor}

\usepackage{placeins}
\usepackage{hyperref}
\usepackage{enumitem}
\usepackage{algorithm}
\usepackage{algpseudocode}
\usepackage[caption=false,font=footnotesize]{subfig}

\begin{document}

\def\BibTeX{{\rm B\kern-.05em{\sc i\kern-.025em b}\kern-.08em
    T\kern-.1667em\lower.7ex\hbox{E}\kern-.125emX}}

\newcommand{\ido}[1]{{\color{orange}Ido: [#1]}}
\newcommand{\marek}[1]{{\color{violet}Marek: [#1]}}
\newcommand{\peter}[1]{{\color{cyan}Peter: [#1]}}

\newcommand{\sulf}{SO\textsubscript{2}F} 

\title{Practical Scalability of Tensor Network Quantum Emulators for Molecular Hamiltonian Simulation}
\author{

\IEEEauthorblockN{Marek Kowalik}
\IEEEauthorblockA{\textit{Capgemini Quantum Lab}\\
Paris, France\\
ORCID: 0009-0007-8484-1315}

\and
\IEEEauthorblockN{Ellen Michael}
\IEEEauthorblockA{\textit{Capgemini UK}\\
London, United Kingdom\\
ORCID: 0009-0002-8750-0376
}

\and
\IEEEauthorblockN{Peter Pog\'{a}ny}
\IEEEauthorblockA{\textit{GSK Medicines Research Centre}\\
Stevenage, United Kingdom\\
ORCID: 0000-0003-3536-0746
}

\and
\IEEEauthorblockN{Phalgun Lolur$^*$}
\IEEEauthorblockA{\textit{Capgemini Quantum Lab}\\
Paris, France\\
Email: phalgun.lolur@capgemini.com\\
$^*$Corresponding author}
}

\maketitle

\begin{abstract}
Quantum computing holds promise for advancing computational chemistry, but near-term quantum hardware remains limited by noise and scale, motivating the use of classical bridging technologies, such as quantum emulators. This work presents an application-specific systems-level benchmark of matrix product state (MPS) tensor-network emulation for real-time Hamiltonian evolution of a density matrix embedding theory (DMET)-embedded sulfonyl-fluoride pharmaceutical fragment, using state-vector simulation as reference. We evaluate runtime scaling, accuracy, resource requirements and entanglement growth across active space sizes ranging from 4 to 24 qubits, for a one-body temporal observable used as a quantum fingerprint for reactivity prediction. The results identify a clear practical boundary for this workflow. At fixed bond dimension, MPS emulation retains the expected favorable scaling, but the bond dimension needed to estimate the target observable within a 1.6 mHa threshold derived from chemical-accuracy requirements grows rapidly with active-space size. In the 20-24 qubit regime, the required bond dimension approaches the maximum available MPS representation, eliminating the runtime advantage over state-vector simulation. Entanglement entropy analysis shows that this cost growth is driven by increasing bipartite entanglement in the time-evolved state, consistent with the mismatch between a one-dimensional MPS ansatz and the non-local correlations generated by molecular electronic dynamics. This study does not claim a universal crossover for all molecules, observables, mappings, or tensor-network geometries. Rather, it provides an application-specific, end-to-end measurement of where a widely used MPS emulation strategy ceases to be an efficient classical surrogate for a chemically motivated Hamiltonian-simulation workflow. The results motivate entanglement-aware algorithm design, orbital-ordering and mapping optimization, alternative tensor-network geometries, and ultimately fault‑tolerant quantum hardware approaches for regimes where accurate molecular dynamics generate non-compressible entanglement.
\end{abstract}


\section{Introduction}
\label{sec:introduction}

Computational chemistry simulations play a central role in modern drug discovery, enabling the prediction of molecular properties, reaction pathways, and chemical reactivity \cite{Sadybekov2023}. However, accurate simulation of large, electronically complex molecules remains computationally demanding for classical computers, particularly when quantum many-body effects become important \cite{Chan2024}. Quantum algorithms for Hamiltonian simulation offer a principled route to address these challenges by directly representing quantum dynamics \cite{Cao2019,Chan2024}. In practice, however, their deployment is constrained by the limited qubit counts, noise, and coherence times of current quantum hardware \cite{Johnstun2021}. 

This gap between algorithmic promise and hardware capability has motivated the development of classical quantum emulation techniques as bridging technologies. Among these, tensor network–based emulators have attracted particular attention due to their ability to compress quantum states by exploiting structure in entanglement \cite{Perez2007,Orus2014,Nguyen2021}. MPS representations, in particular, have been widely adopted because of their conceptual simplicity, numerical robustness, and scalability for systems with limited or structured entanglement. Recent work has demonstrated the effectiveness of MPS emulators for certain highly structured quantum circuits, in some cases extending to hundreds or even thousands of qubits \cite{leonteva2024comparative,Schieffer2025}. 

Whether such favorable scaling persists for chemically realistic Hamiltonian simulation workloads remains an open systems-level question. Molecular electronic structure problems involve non-local interactions, dense Hamiltonians, and time-evolved quantum states that are not generally expected to obey one-dimensional area laws. These characteristics raise concerns about whether MPS-based emulators can maintain accuracy without incurring rapidly increasing bond dimensions, undermining their practical runtime advantages. Understanding *where* and *why* such emulators cease to be efficient is essential for determining their role in end-to-end quantum chemistry workflows.

We emphasize that the qualitative mechanism behind MPS breakdown is not claimed to be new: one-dimensional MPS representations are known to be efficient primarily when the relevant entanglement remains sufficiently bounded, as in many one-dimensional local systems satisfying area-law behavior \cite{Orus2014,Schuch_2008,Brand_o_2013, Eisert_2010}. The contribution of this work is instead empirical and systems-level: we quantify the bond-dimension, runtime, and memory consequences of this mechanism in an end-to-end molecular Hamiltonian-simulation workflow motivated by pharmaceutical reactivity prediction.

Targeted covalent drugs provide a stringent and practically relevant stress test for these questions \cite{Boike2022,Sutanto2020,Singh2022}. Predicting their reactivity requires resolving subtle electronic effects associated with bond formation and charge redistribution, often under tight accuracy constraints \cite{AWOONORWILLIAMS2021203,Zhang2025}. Recent work has shown that time-dependent quantum observables derived from Hamiltonian simulation can serve as informative features for machine-learning–based reactivity prediction \cite{Montgomery2023}. At the same time, these workflows demand simulation of molecular fragments at active space sizes that quickly push beyond the regime where classical state-vector simulation remains tractable.

This work addresses the following systems-level question: \textit{do MPS-based tensor network quantum emulators remain computationally viable at the active space sizes where molecular Hamiltonian simulations become chemically meaningful?} To answer this for a concrete chemically motivated workflow, we benchmark MPS emulation of real-time Hamiltonian evolution for one DMET-embedded pharmaceutical fragment. We evaluate runtime scaling, accuracy, and resource requirements across active space sizes from 4 to 24 qubits, using state-vector simulation as ground truth.

MPS-based emulators were selected due to their widespread use and their ability to provide direct insight into the entanglement structure of the simulated quantum states. Other tensor network geometries are left for future comparison. By quantifying the bond dimension required to achieve chemically accurate estimates of a relevant temporal observable, and by relating this requirement to measured entanglement entropy, we identify the regimes where MPS compression is effective and where it fundamentally breaks down. The results provide a concrete benchmark of this workflow and a protocol for testing whether similar MPS limitations arise in other molecular systems. 

\section{Methods}
\label{sec:methods}

The methodology focuses on three core aspects: emulator configuration, molecular system simulation, and accuracy benchmarking. We aim to understand how emulator performance scales with system size and to identify the practical limitations of tensor network methods in capturing quantum features critical for drug discovery. The final part of this section proposes criteria for evaluating emulator efficiency, which serve as a foundation for interpreting the results and discussing the broader implications for quantum computing in chemistry.

\subsection{Benchmarking Protocol}
To ensure reproducible and rigorous evaluation of tensor network emulators, we implemented the following benchmarking protocol:

\begin{enumerate}
    \item \textbf{Emulator configuration:} Ava emulator developed by Fermioniq, a state-based, MPS tensor network emulator with configurable bond dimensions. The emulator's quantum circuit transpilation pipeline included a pass through Qiskit's transpiler using the \(Clifford+R_x+R_y+R_z\) gate basis with default settings, followed by Ava's built-in transpilation routine. No SWAP-network or gate-order optimization was applied to minimize intermediate MPS bond growth. Only this MPS emulator implementation was benchmarked, so runtime and numerical behavior are implementation-specific.
    \item \textbf{Ground truth:} Ideal time evolution trajectories, up to 24 qubits, were calculated using state-vector simulation.
    \item \textbf{Performance metrics:}
    \begin{itemize}
        \item \textit{Runtime scaling} - Measured as wall-clock time for complete simulation execution across different active space sizes
        \item \textit{Accuracy} - Defined as deviation, measured as mean absolute error, from ground truth expectation values obtained via state-vector simulation
        \item \textit{Resource requirements} - Memory usage and computational resources needed for simulation
    \end{itemize}
    
    \item \textbf{Validation approach:} For each active space size, we varied the bond dimension and compared the estimated value of the target observable \(F(t)\) at \(t=10\) a.u. to state-vector ground truth. The benchmark is therefore a task-level observable-estimation test for the quantum-fingerprint workflow, not a certification of global state fidelity or of accuracy for arbitrary molecular observables. Following established computational chemistry standards \cite{Lolur2023, Cao2019, Korhonen2024}, we defined 'accurate' results as those falling within 1.6 mHa ($\sim$1 kcal/mol or 4.18 kJ/mol) of the ground truth.
    
    \item \textbf{Data analysis:} To provide insights into the practical utility and fundamental limitations of MPS emulators for this use case, for all collected runs: 
    smallest bond dimensions yielding \(F(t)\) within defined threshold, and their respective runs (across active spaces) was extracted.
\end{enumerate}

Importantly, no orbital-reordering or qubit-mapping optimization was applied; all active spaces used the fixed orbital order and Jordan--Wigner mapping of the production workflow, so the reported bond dimensions should be interpreted as workflow-specific rather than ordering-optimized.

\subsection{Molecular System and Simulation Protocol}
\label{sub_sec:molecular_system_and_simulation_protocol}

We selected a representative molecule (4-(Morpholine-4-sulfonyl)benzene-1-sulfonyl fluoride) from a series of targeted covalent drugs with reactive sulfonyl fluoride fragment $SO_2F$ , that was embedded using the DMET\cite{Wouters_2016} approach. This molecule should be viewed as a chemically motivated test case rather than a statistically representative sample of pharmaceutical electronic structures. Sulfonyl fluorides are crucial in medicinal chemistry and drug development due to their ability to act as versatile electrophiles, facilitating the formation of covalent bonds with nucleophiles and enabling the design of potent inhibitors for various biological targets. Molecular data and end-to-end data-driven approach are taken from \cite{Montgomery2023} - they are described there broader, along with rest of the use case details. 
For this molecule, we simulated real-time evolution across a range of active space sizes, from 4 to 24 spin orbitals (corresponding to 4-24 qubits after Jordan--Wigner qubit mapping), with the number of electrons being equal to half of the number of spin orbitals. Because Jordan--Wigner maps each spin orbital occupation mode to one qubit, 4-24 spin orbitals correspond to 4-24 qubits, i.e. 2-12 spatial orbitals with both spin projections included. This range covers most pharmaceutically relevant scenarios\cite{Izs_k_2022}, where organic molecules typically involve p and s orbitals, with up to 8 electrons per atom. While larger active spaces (up to 24-26 electrons) might be relevant for systems containing f-elements or for specialized cases involving d-orbitals, such scenarios are rare in pharmaceutical applications. Considering runtimes of quantum computers on those active space sizes (with and without quantum error corrections overheads), robustness of classical computation methods in this regime, and the expected runtime advantage in larger active spaces for quantum computers, the given use case should also be extended with an evaluation of the reactivity prediction using larger active space sizes to potentially justify using larger active spaces, based on the prediction accuracy improvement. For the scope of this work, the authors limit the analysis to the active spaces mentioned before.

The effective electronic structure Hamiltonian of the embedded fragment was constructed for each active space size using PySCF\cite{Sun2018} and Vayesta\cite{Nusspickel2023}, following approach from \cite{Montgomery2023}. The time evolution simulations were performed using a first-order Trotter decomposition\cite{Lloyd1996,Berry2006} scheme. The molecular geometries were taken from the dataset from \cite{Montgomery2023}.  

The final goal of the quantum algorithm there is to provide the input feature (\textit{`quantum fingerprint'}\cite{Montgomery2023}) for a machine learning model to accurately predict the reactivity of the molecules. In the most naive approach, one could use as a feature the energy trajectory sampled over a set of time steps. Following the original manuscript findings\cite{Montgomery2023}, in this work, the input feature would be indeed the trajectory sampled over a range of time steps, yet the measurement operator would be single temporal observable:

\begin{equation}
    F(t) = \sum_{r,s} h^{eff}_{rs} \rho_{rs}(t) 
    \label{eq:quantum_fingerprint} 
\end{equation}

Here, $\rho_{rs}(t)$ denotes the elements of the one-body density matrix of the embedded fragment active space (so that the number of terms in the observable sum scales quadratically with the active space size):

\begin{equation} 
    \rho_{rs}(t) = \langle \psi(t)| \hat{a}_r^{\dagger} \hat{a}_s|\psi(t)\rangle . 
    \label{eq:one_rdm}    
\end{equation}

and $h^{eff}_{rs}$ represents the effective single-electron integrals as defined in~\cite{Montgomery2023}, with \(r\) and \(s\) indexing localized spin orbitals.

The emulation is performed at \(t=10\) a.u. using one first-order Trotter step, matching the benchmark observable in the source workflow \cite{Montgomery2023}. There, using more Trotter steps did not materially change the target observable in the considered regime. We nevertheless treat step count as a fixed resource parameter: additional steps or higher-order formulas would increase circuit depth and runtime, and may change truncation-error accumulation, although they approximate the same fixed-time evolution \cite{Lloyd1996,Berry2006}. It is also important to note that while used time frame is suitable for many electronic transitions, certain processes, particularly those involving spin state transitions, may require significantly longer evolution times, by orders of magnitude\cite{viswanatha2025impactcooperativityspatialtemporal}.

\subsection{Criteria of efficient tensor network emulators} 

The choice of the evaluation criteria for the efficacy of tensor network quantum emulators is cornerstone of the algorithmic design, deciding then on the applicability of emulators. In this subsection, we propose a set of criteria for evaluating the computational efficiency of TN-emulators.

\begin{itemize}
    \item \textbf{Scalability} - low-degree polynomial scaling of runtime for fixed quantum circuit depth, prohibiting the explosion of the emulation resources.

    \item \textbf{Accuracy} - precision of the emulation results up to the computational accuracy required by the problem or arbitrarily defined to accept the problem as solved accurately enough.
    \item \textbf{Efficient resource utilization } - proper utilization of the classical computational backends \cite{Patra_2024,Tindall_2024} in particular for GPU accelerated implementations\cite{pan2024} and in general on heterogeneous backends. 
    \item \textbf{Reliability. } -  stable performance, with similar runtime and results for the same emulation problem for different trials. Emulation may contain some random elements (e.g. recomposition of the circuit, choosing the contraction path), yet the runtimes and the results should not vary strongly or often. One remark to this criterion is that emulator might be extended with the functionalities to early detect slow or inaccurate runs and restart the run.
\end{itemize}

The proposed criteria need to be evaluated for a particular use case through its computational problem with a concrete quantum algorithm (e.g. Quantum Phase Estimation (QPE) \cite{kitaev1995}), its concrete implementation (e.g. Iterative QPE from \cite{Dob_ek_2007}) for particular quantum backend type (e.g. gate-based backends) and its instance for QPU comparison (e.g. IBM Quantum Heron H2 devices \cite{ibm_heron_2024}). Fulfilling them can lead to the scalability of the emulation towards the scales of the classically challenging problems that are out of the state-vector emulation feasible scope. 

\section{Results}
\label{sec:results}

The results are organized into six main categories: runtime scaling analysis, accuracy dependence on bond dimension, circuit complexity considerations, resource requirements, comparative analysis of accuracy and efficiency, and entanglement structure examination. Of particular significance is the relationship between bond dimension requirements and system size, which proves to be a critical factor in determining the viability of these methods for drug discovery applications.

\subsection{Runtime Scaling}
\label{sub_sec:runtime_scaling}

\begin{figure}
    \centering
    \subfloat[]{%
        \includegraphics[width=0.8\columnwidth]{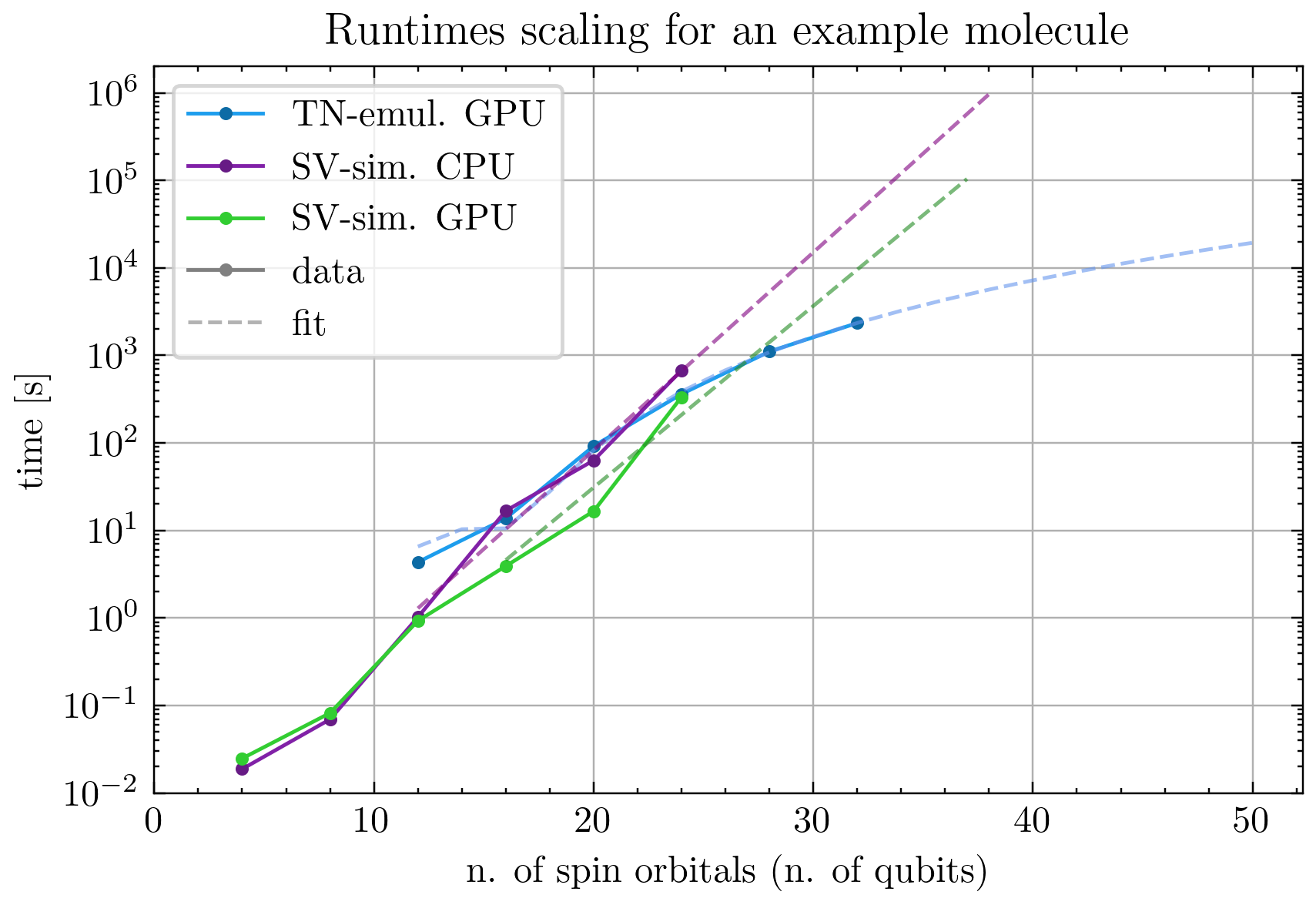}
    }\hfill
    \subfloat[]{%
        \includegraphics[width=0.8\columnwidth]{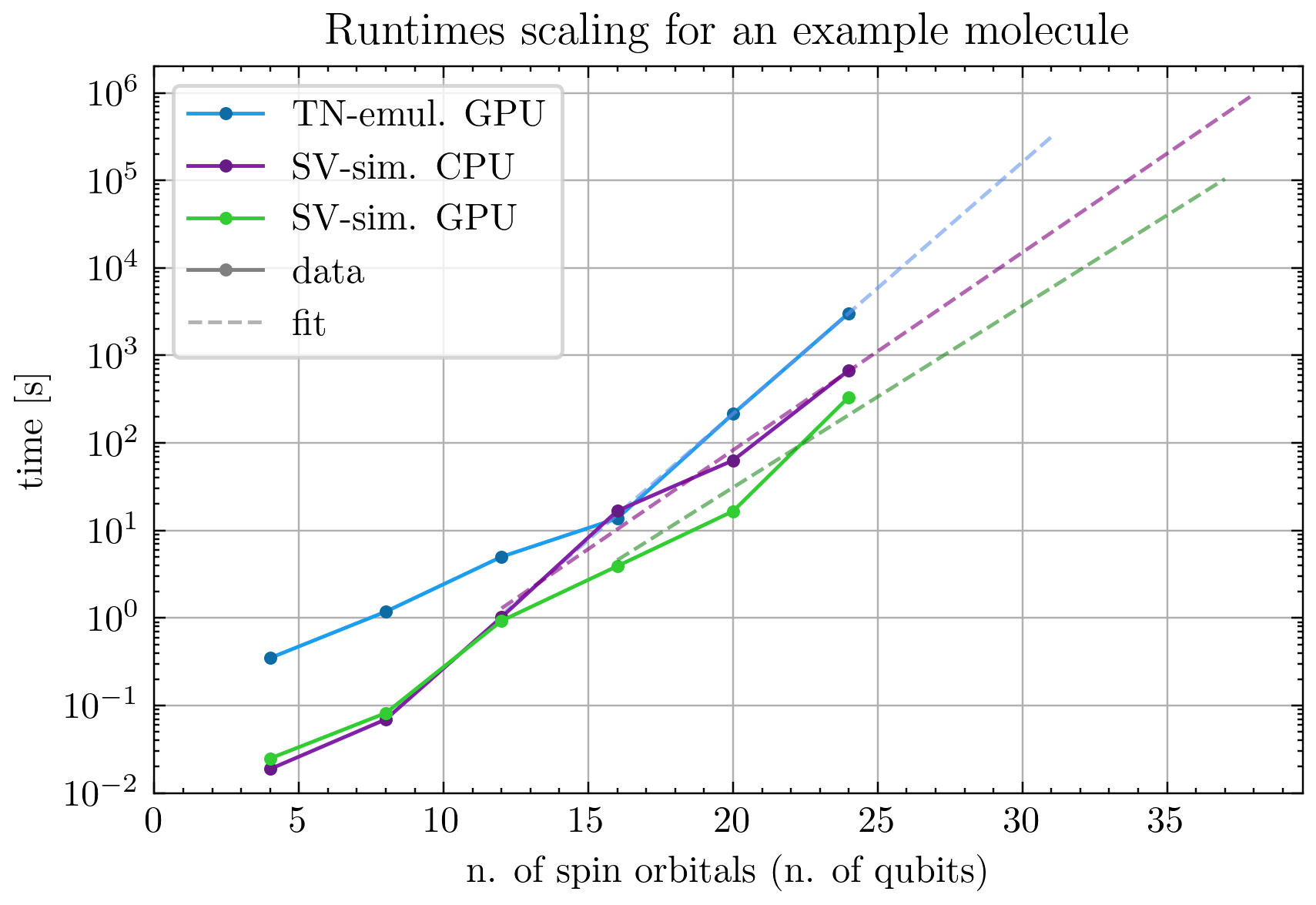}
    }
    \caption[Runtimes scaling for real-time evolution simulations]{%
    Runtimes scaling for the range of active space sizes for real-time evolution simulation for 1 Trotter step on the Qiskit state-vector simulator (`SV-sim.' label) run on CPU and 12GB VRAM GPU and Fermioniq emulator (`TN-emul.' label). Panel a) shows run with bond dimension 1024 on 12GB VRAM GPU, Panel b) - bond dimensions from the Fig.~\ref{fig:suff_bd_from_n_qubits}, on 48GB VRAM GPU. All results were extrapolated with expected fitting functions (exponential for the state-vector simulator, cubic (Panel a)) and exponential on the last 3 points (Panel b)) for the Fermioniq’s emulator). Fits are included only as visual trend guides. Extrapolations beyond the measured 4-24 qubit range, carry substantial uncertainty and are not used to support the paper's main conclusions. The extrapolation limits were based on memory limits and the size of non-sparse arrays of 128-bits complex numbers. For the state-vector simulation on a single CPU, the limit of 4 TB RAM memory per node was assumed, based on the High-Performance Computing centers and supercomputers offering \cite{ddbj2025hardware, eurohpc2025guidelines}, limiting the extrapolation to 38 qubits. For the same state-vector simulator on a GPU, based on the record-breaking GigaIO SuperNODE configuration with 2 TB of VRAM shared within 32 GPUs \cite{gigaio2025supernode}, the limit is set at 37 qubits. For the emulator on the same computational node, the limit of qubits is around 60 000 qubits to express MPS with bond dimension 1024, for max bond dimension, assumed for the bottom panel, it is 31.}
    \label{fig:time_evo_runtime_scaling}
\end{figure}

Fig.~\ref{fig:time_evo_runtime_scaling} illustrates the comparison of runtime scaling between state-vector simulation and tensor network emulation. The first panel a) displays the runtimes of the emulators, for experiments run for fixed bond dimension $D=1024$. This showcases the polynomial scaling of such simulations for fixed bond dimension, under the assumption of fixed or polynomial scaling of circuit depths from active space size, that seems to occurs in our circuits (\ref{fig:circ_depths_from_n_qubits}). The results confirm such scaling. The fitted curves - with cubic function -  in Fig.~\ref{fig:time_evo_runtime_scaling} are intended only as visual guides to compare trends over the measured range and should not be interpreted as reliable runtime forecasts beyond the largest simulated active space.

The second panel in Fig.~\ref{fig:time_evo_runtime_scaling} presents runtime scaling for the minimum bond dimension, allowing to accurately emulate trajectories of given active space. The computational accuracy criterion used there is defined in Subsection \ref{sub_sec:accuracy_vs_bond_dimension} below. The values of the bond dimension used for these simulations are plotted in Fig.~\ref{fig:suff_bd_from_n_qubits}. The scaling behavior exhibits distinct regimes, depending on the system size. For systems below 16 qubits, MPS emulation can achieve the target-observable accuracy with relatively modest bond dimensions, but it does not provide a wall-clock advantage over state-vector simulation in this measured range. Its potential benefit in this regime is therefore not faster runtime, but the lower compressed-state representation at fixed bond dimension and the ability to expose entanglement-driven scaling trends. However, this advantageous scaling diminishes rapidly for larger systems, particularly when higher bond dimensions are required for accuracy - for 16+ qubits. For the fixed orbital ordering and qubit mapping used here, the crossover point where MPS emulation loses its runtime advantage appears around 20-24 qubits. This behavior correlates strongly with the increasing entanglement in the molecular systems, requiring larger bond dimensions for the accurate representation of the quantum state.  

\noindent

\subsection{Accuracy vs Bond Dimension}
\label{sub_sec:accuracy_vs_bond_dimension}

\begin{figure}
    \centering
    \subfloat[]{%
        \includegraphics[width=0.78\columnwidth]{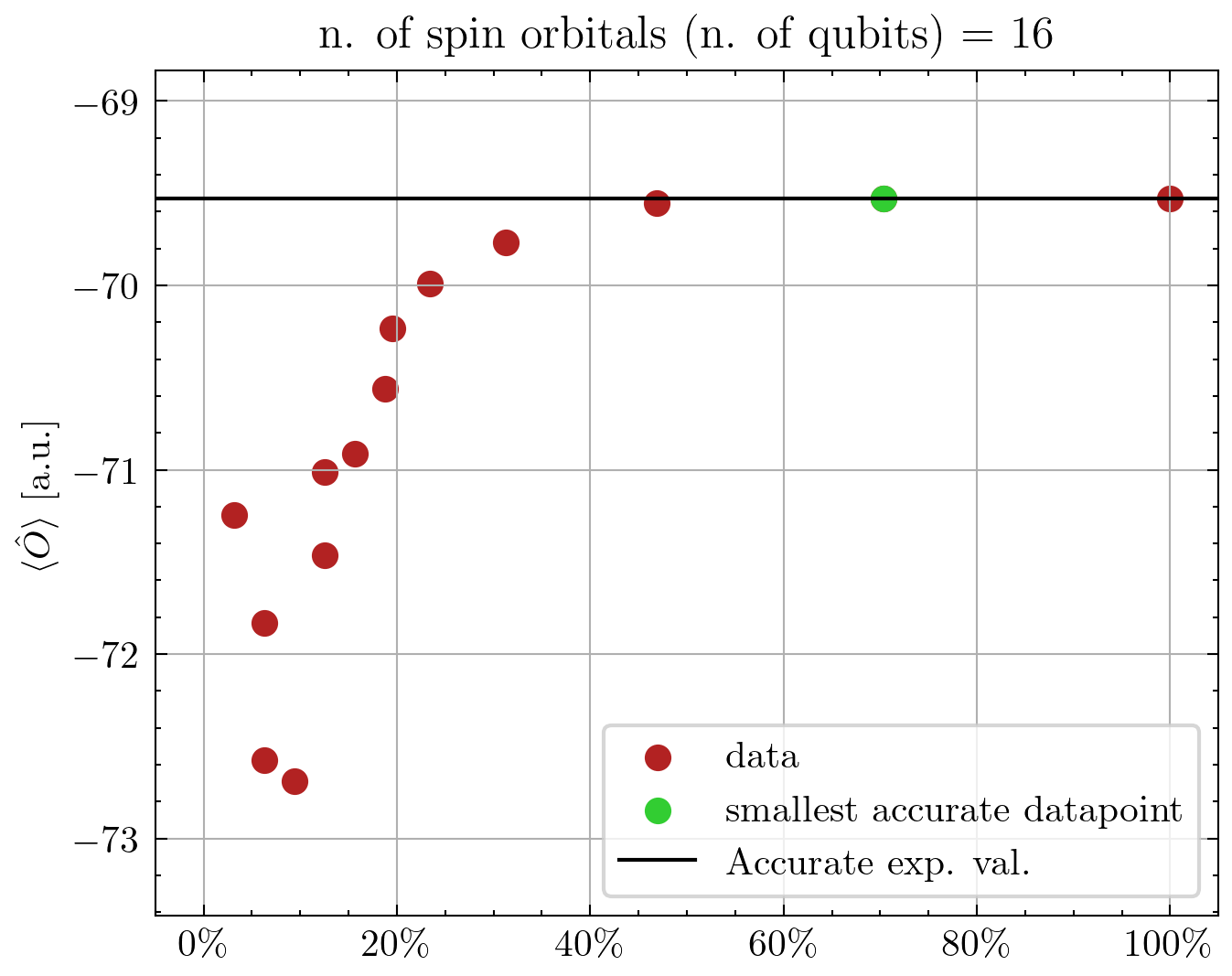}
    }\hfill
    \subfloat[]{%
        \includegraphics[width=0.78\columnwidth]{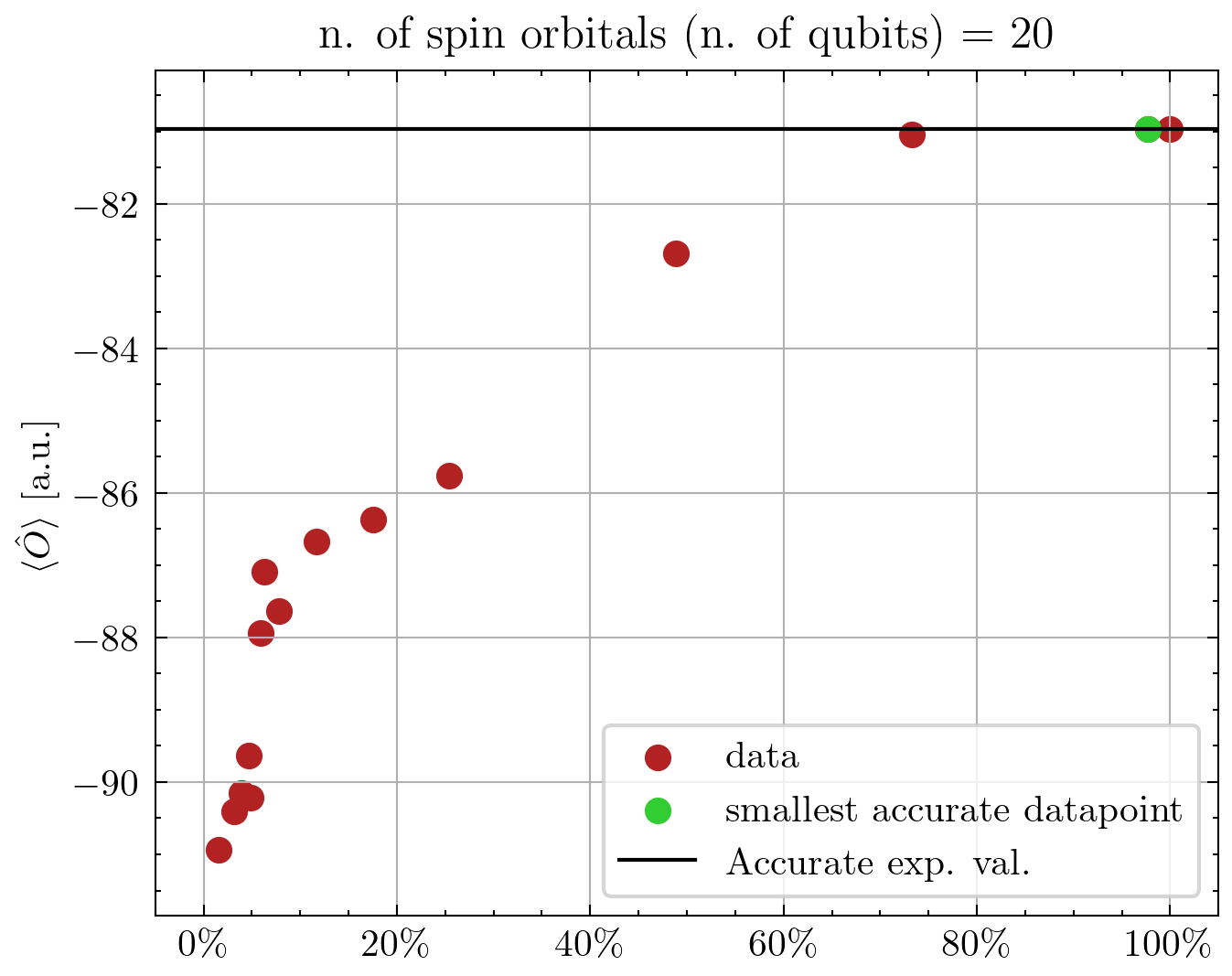}
    }
    \caption[]{The expectation values of the temporal observable estimated with emulation with different bond dimension values for an exemplary molecule with active spaces of (a) 8 electrons, 16 spin orbitals (16 qubits) and (b) 10 electrons, 20 spin orbitals (20 qubits). The bond dimensions were converted into percentages of the maximum bond dimension for the emulated number of qubits for better readability. The data point with the smallest bond dimension that yielded computationally accurate results is marked with a green dot per each active space.}
    \label{fig:accuracy_bd_trade_off}
\end{figure}

Fig.~\ref{fig:accuracy_bd_trade_off} demonstrates the convergence of estimated expectation values with increasing bond dimension for examples of 16-qubit and 20-qubit systems. This plot reveals the non-trivial relationship between bond dimension and accuracy, highlighting the challenge of selecting appropriate approximation levels.

\noindent

\begin{figure}
    \centering
    \includegraphics[width=0.8\linewidth]{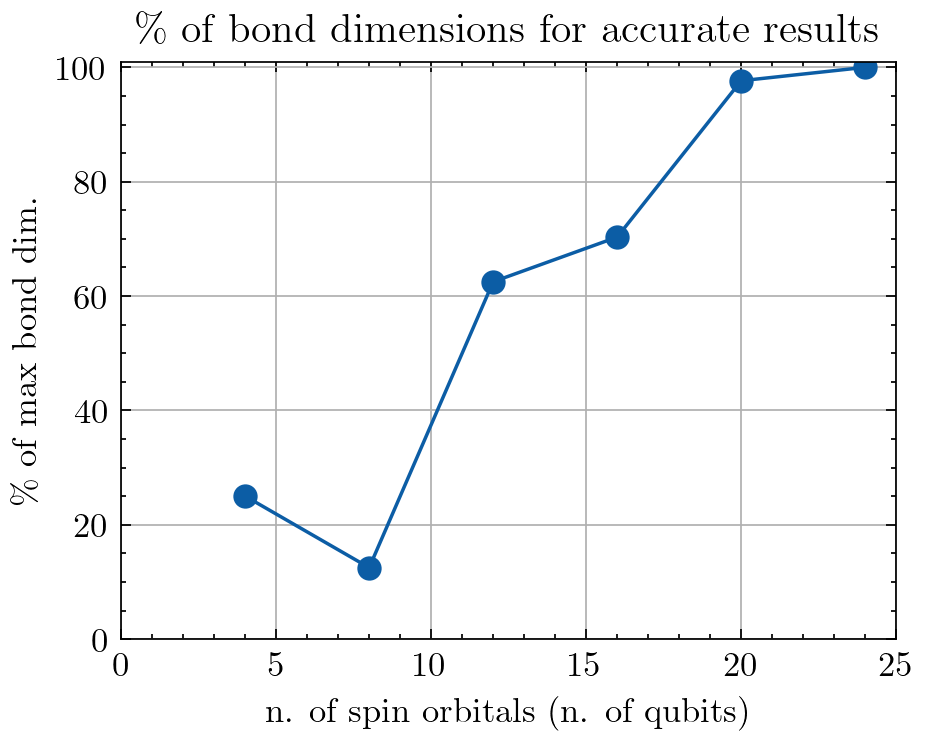}
    \caption{The smallest bond dimensions (shown as a percentage of maximum bond dimension for a given number of qubits) for which the emulation yielded the estimated observable expectation value error within the defined accuracy threshold for a range of active space sizes (spin orbitals, hence the number of qubits).}
    \label{fig:suff_bd_from_n_qubits}
\end{figure}

Fig.~\ref{fig:suff_bd_from_n_qubits} extends this analysis across all active spaces, showing the minimum bond dimension required to estimate the target observable within 1.6 mHa of state-vector ground truth. Other observables, especially higher-body correlators or phase-sensitive quantities, may require different bond dimensions and are not benchmarked here. A clear trend emerges: as the active space size increases, the required bond dimension grows rapidly, approaching the maximum possible value for larger systems. 

For the 24-qubit active space studied here, chemical accuracy-derived precision required the maximum bond dimension accessible in the MPS representation. Within this benchmark, this eliminates the practical runtime advantage over state-vector simulation at the largest active space tested, and if the runtime would follow the exponential runtime scaling trend of emulators, illustrated in panel b) of Fig.~\ref{fig:time_evo_runtime_scaling}, also all the larger active spaces for the tested system. We therefore avoid assigning a quantitative failure point beyond 24 qubits; the measured data already show that, at the largest active space tested, the bond dimension required for accuracy removes the practical MPS advantage in this workflow. The relationship between accuracy and bond dimension shown here is the indicator of the non-1D entanglement structure of the system, particularly for systems larger than 16 qubits. We also observed that expectation value estimates initially converge rapidly with increasing bond dimension, followed by a plateau region where substantial increases in bond dimension yield only marginal improvements in accuracy. This behavior suggests a practical limitation of the MPS representation for this molecular workflow. Starting for the most efficient to emulate accurately (8-qubit system), achieving the 1.6 mHa observable-error threshold required bond dimensions around 12.5\% of the maximum possible value, then start to rise rapidly, reaching 100\% for 24-qubits,  indicating significant entanglement, that is non-1D in its structure, in the simulated quantum state that resists efficient compression and raise the cost of the MPS emulation to maximal one, yet before reaching active space sizes challenging for classical simulation.

\subsection{Circuit Complexity}
\label{sub_sec:circuit_complexity}

\begin{figure}
    \centering
    \subfloat[]{%
        \includegraphics[width=0.8\columnwidth]{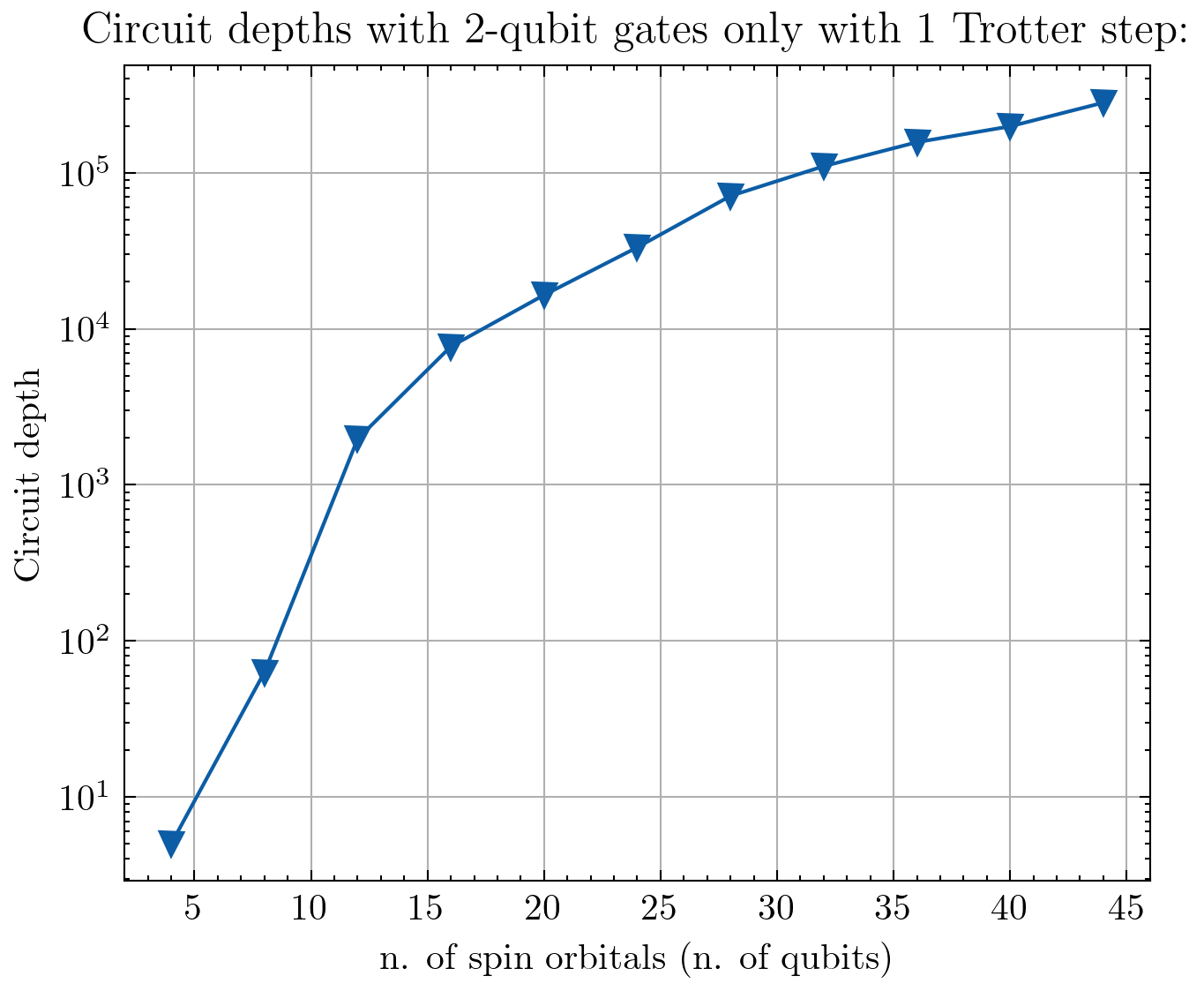}
    }\hfill
    \subfloat[]{%
        \includegraphics[width=0.8\columnwidth]{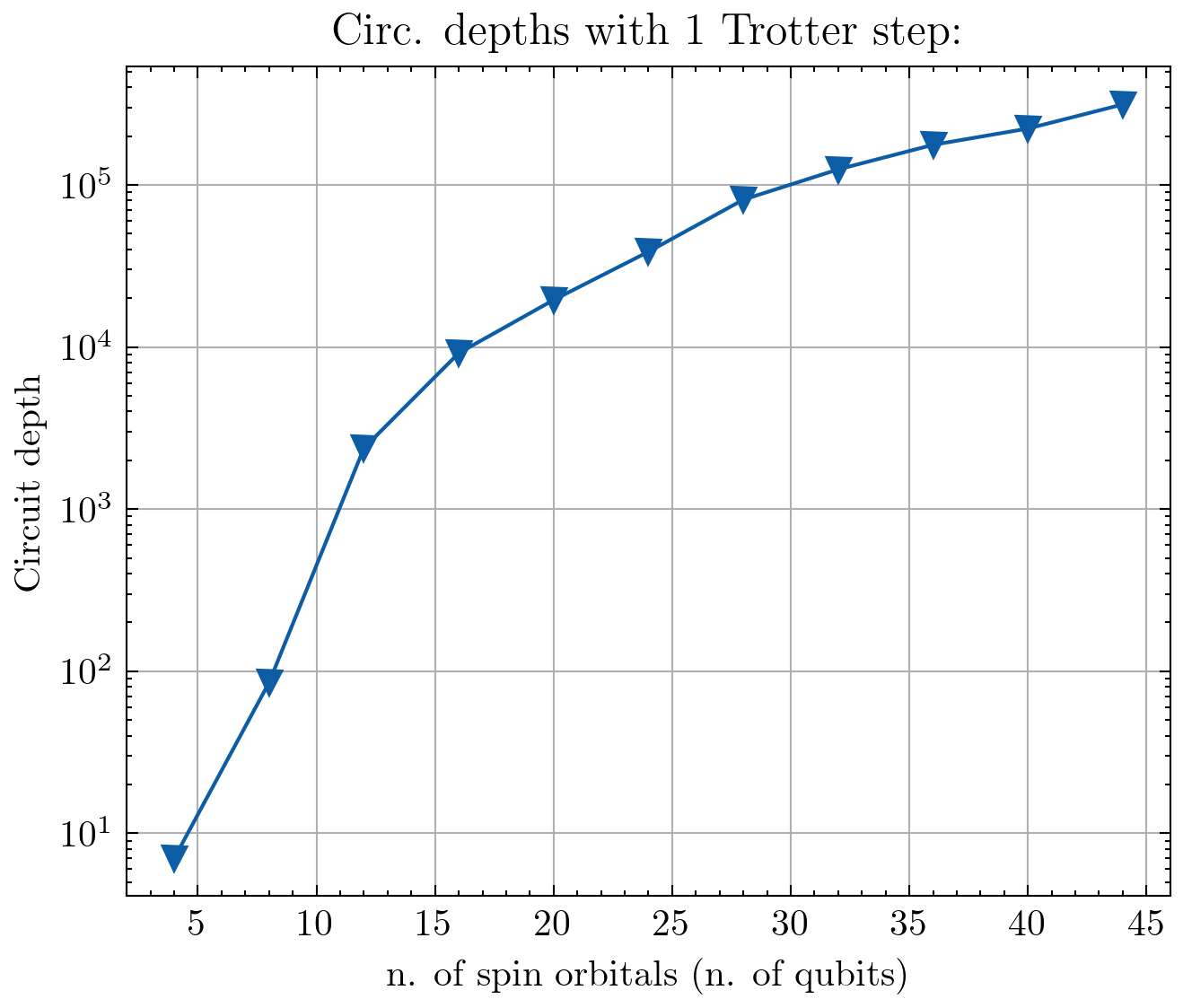}
    }
    \caption[]{Quantum circuit depths of the time evolution circuit emulation for a range of active space sizes (hence number of qubits). The depth from circuit with all gates included is displayed in panel (a) and the depth of only 2-qubit gates counted is in panel (b). After synthesis using Jordan--Wigner qubit mapping and first order product formula decomposition\cite{Lloyd1996,Berry2006} with 1 Trotter step in Qiskit\cite{Qiskit}, the circuits were transpiled to basis set of gates 
    \{$RXX,RYY,RZX,RZZ,CX,XX-YY,XX+YY,H,RX,RY,RZ$\} into all-to-all connectivity simulator backend with optimization level 2. }
    \label{fig:circ_depths_from_n_qubits}
\end{figure}

An additional challenge is revealed in Fig.~\ref{fig:circ_depths_from_n_qubits}, which shows that the depth of the circuit increases substantially with the active space size, following increasing complexity of the molecular Hamiltonian terms. This growth in circuit complexity further challenges efficient emulation, as deeper circuits generally lead to more entangled states that are harder to approximate with tensor networks. Additionally, sole number of qubit-mapped Hamiltonian terms, i.e. Pauli terms, scales polynomially with the active space size\cite{Chan2024}, contributing significantly to the overall computational cost. Deeper circuits also lead to an increase in the accumulation of numerical errors in the tensor network representation, requiring higher bond dimensions to maintain accuracy. This compounding effect creates a particularly challenging computational scenario for larger molecular systems, where both the circuit depth and the required bond dimension grow substantially.

\noindent

\subsection{Resource Requirements} \label{sub_sec:resource_requirements}

The computational resource requirements for accurate MPS emulation show steep scaling with system size. Memory usage, in particular, becomes a limiting factor for larger systems due to the need for increased bond dimensions. Following the MPS size, the memory requirements grow approximately as $O(D^2N)$ where $D$ is the bond dimension and $N$ is the number of qubits. However, this low-degree polynomial scaling becomes irrelevant when the required $D$ to reach the 1.6 mHa observable-error threshold in the simulation scales seemingly exponentially in system size. For systems beyond 20 qubits, even with high-end GPU resources, memory constraints begin to limit the maximum achievable bond dimension, directly impacting simulation accuracy. These resource limitations are relevant for workflows requiring chemically meaningful observable precision. More details on the resources utilized in the experiments are in Appendix \ref{app_sec:emulation_run_details}.

\subsection{Comparative Analysis of Accuracy and Efficiency}

\noindent
\begin{figure}
    \centering
    \includegraphics[width=0.85\linewidth]{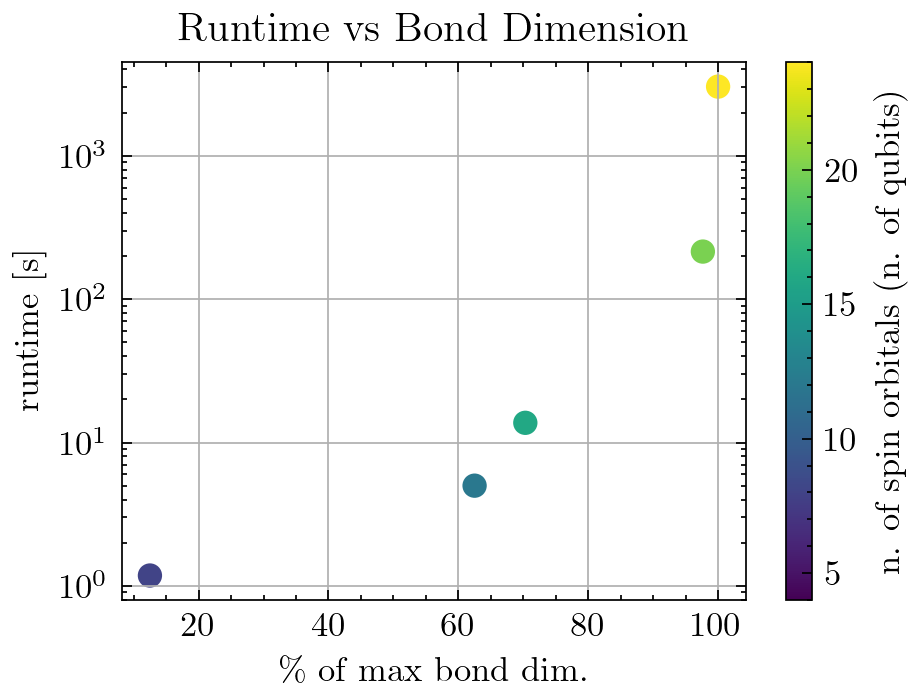}
    \caption{The runtimes over the active space sizes runs (with smallest bond dimensions for accurate target-observable estimation) plotted from their respective bond dimensions (shown as a percentage of maximum bond dimension for a given number of qubits). 4-qubits case, as an outlier, was excluded.}
    \label{fig:comparative_analysis}
\end{figure}

To better understand the relationship between accuracy, bond dimension, and computational efficiency, we conducted a comparative analysis across all active space sizes. Fig.~\ref{fig:comparative_analysis} presents a visualization of this three-way relationship, plotting the minimum bond dimension required to achieve desired accuracy (within 1.6 mHa of ground truth) against runtime implications through active space size.

Our analysis reveals a critical inflection point at approximately 12 qubits, where the increasing normalized bond dimension substantially increases runtime, even for the logarithmic scale. For systems below this threshold, MPS emulation can achieve the target observable accuracy with relatively modest bond dimensions (10-15\% of maximum), offering potential computational advantages over state-vector methods despite some overhead costs. However, for systems of 16 qubits and above, the required bond dimension, grows rapidly, approaching 100\% of the maximum possible value for 24-qubit systems, hence runtime increases sharply once the 1.6 mHa observable-error threshold constraints are enforced.

This finding has significant implications for the practical utility of MPS-based emulation in pharmaceutical applications. While MPS methods show promise for smaller molecular fragments, their efficiency advantage diminishes at the upper end of the active-space range considered for this pharmaceutical-fragment benchmark. The exponential growth in required bond dimension effectively negates the polynomial runtime scaling advantage that tensor networks theoretically offer, resulting in computational costs that approach or exceed those of direct state-vector simulation for chemically relevant system sizes.

Furthermore, our comparative analysis highlights that the crossover point where simple MPS emulation loses its advantage coincides with the emergence of complex entanglement patterns in the molecular systems that resist efficient representation in a one-dimensional MPS structure. This observation aligns with theoretical expectations regarding the limitations of MPS representations for systems with non-local interactions and multidimensional entanglement structures.

\subsection{Entanglement structure examination} 
\label{sub_sec:entanglement}

Bond dimension and consequently the resources needed to accurately represent the quantum state depend on the entanglement present in an emulated system \cite{Schuch_2008}. One of the metrics quantifying the level of entanglement is the bipartite entanglement, in particular Von Neumann entanglement entropy \cite{Brand_o_2013} as defined in \cite{Horodecki_2009,Bennett_1996}. 

We kept the orbital ordering and Jordan--Wigner mapping fixed across all experiments. This gives a consistent production-style benchmark, but not an ordering-optimized MPS calculation; orbital ordering is known to affect DMRG/MPS convergence and bond-dimension requirements in quantum chemistry \cite{Chan2011DMRG,Barcza2011QuantumInfo,Szalay2015}.

The MPS representation of a quantum state decomposed into two subsystems with bond dimension $D$, bounds the bipartite entanglement $S$ of the system it can express as $S < log_2(D)$ \cite{Pollmann2015EfficientNS,Preisser_2023}.  

That being said, the larger the entanglement of any subpart of the emulated quantum system, the larger bond dimension required to express it accurately. In particular, if the largest bipartite entanglement grows linearly with the emulated system size, one can expect the required bond dimension to grow exponentially. If the entanglement saturates to some value at given system size range, one can expect the required bond dimension also to saturate. In this scenario, the complexity of the emulation runtime with MPS of such system should be polynomial from number of qubits, assuming polynomial scaling of the circuit depth from number of qubits. Such phenomena occurs in so-called area law for 1D system, and it is fulfilled for the ground state of a local, gapped Hamiltonians. 
However, the non-local interactions and the time-evolved states are in general not expected to follow 1D area law. For this paper's case of the presented electronic structure (qubit mapped and represented with MPS), the scaling of the entanglement rise  was checked empirically through the emulation and measuring the entanglement entropy scaling through the active space sizes.

\begin{figure}
    \centering
    \includegraphics[width=0.8\linewidth]{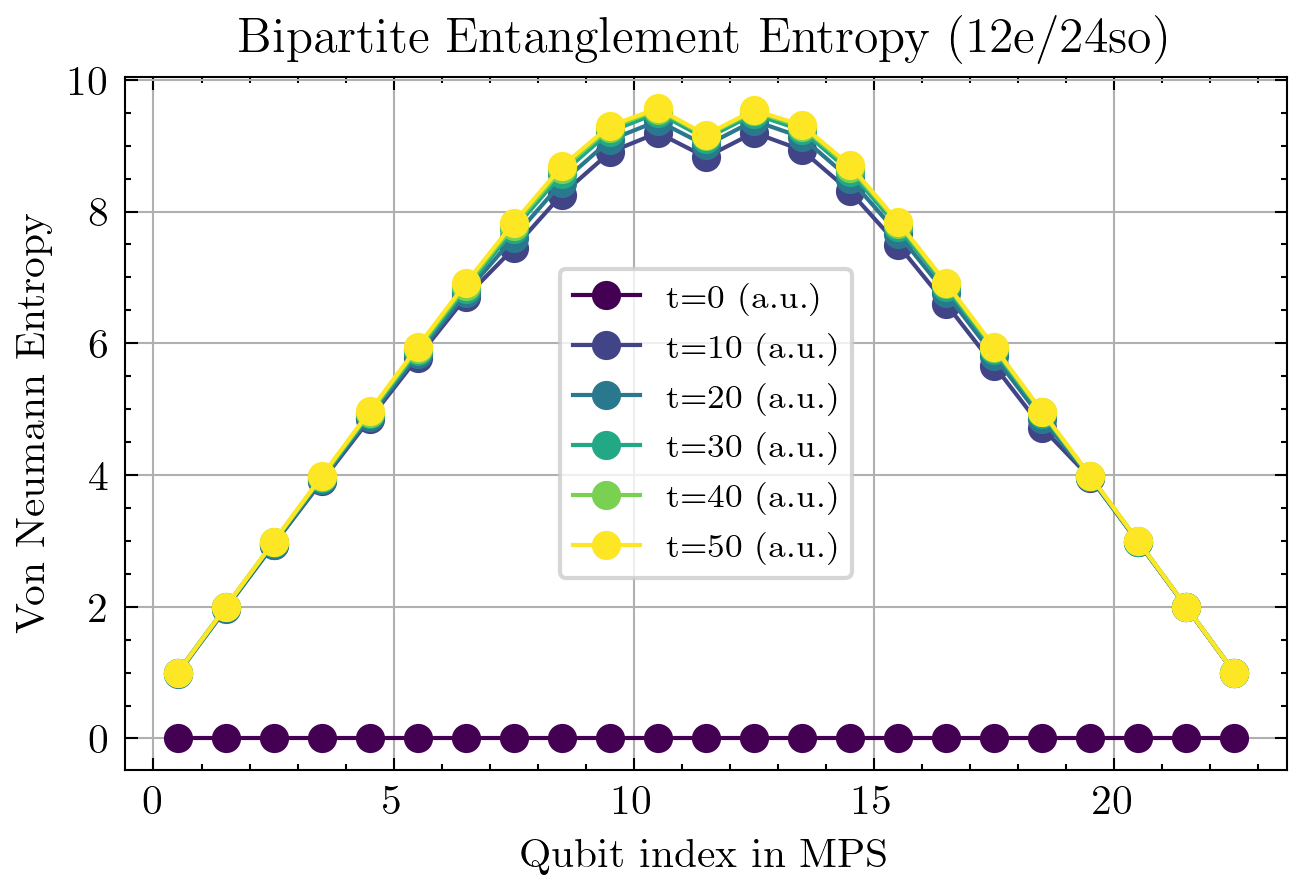}
    \caption{Von Neumann entropy for bipartite splits along the direct qubit connections in MPS for 24 qubits active space example for a range of time steps from 0 to 50 atomic units (a.u.).}
    \label{fig:bip_ent_entropy_24q}
\end{figure}

Measuring naively the bipartite entanglement in the qubit-mapped simulated electronic structure system or, in general, the entanglement for a state of a set of qubits, is a challenging task. The number of all combinations of bipartite split grows exponentially (one can calculate the exact number for $k$ qubits from the sum of binomial coefficients $\sum_{n=1}^{\lceil k/2 \rceil} \binom{k}{n}$), due to the combinatorial nature of the problem. However, for MPS structure, all the qubits are connected in a 1D chain, so to estimate the sufficient bond dimension for accurate quantum state representation, one needs to check the entanglement for each bipartite division of the state vector, as these divisions are interconnected through the indices in the MPS, connecting the tensors representing each qubit. Examples of such entropy values calculated from the ideal simulation of the state vector without sampling on the largest active space tested are shown in Fig.~\ref{fig:bip_ent_entropy_24q}. Two main observations for this active space size are the following:

\begin{enumerate}
    \item the entanglement rises for the division with more and more qubits in smaller part until it slightly drops for the equal split
    \item there is no entanglement in \textit{time=0}, as expected, and there is barely any difference between the rest of the time steps in the entanglement

\end{enumerate}

\noindent

\noindent
\begin{figure}
    \centering
    \subfloat[]{%
        \includegraphics[width=0.85\columnwidth]{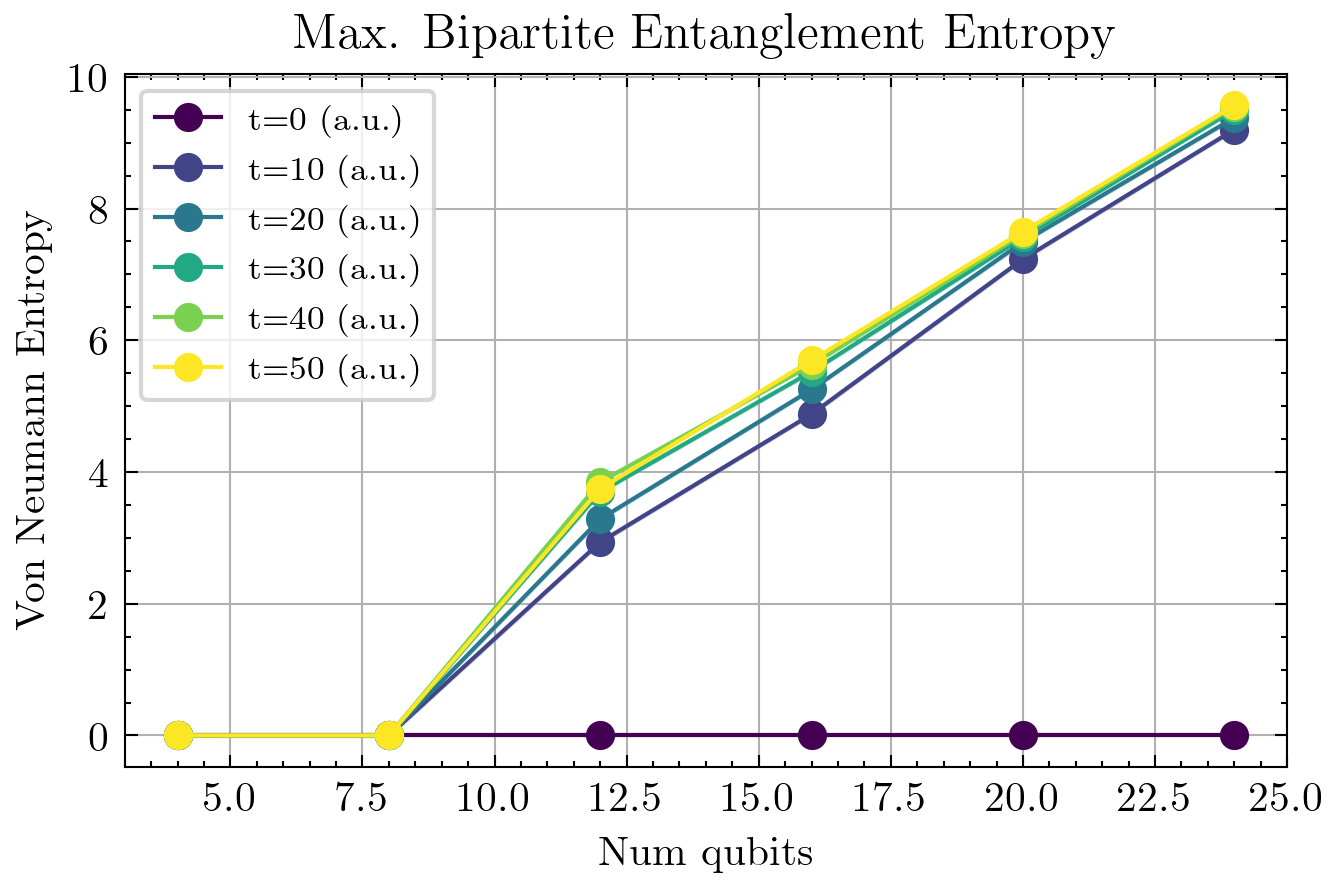}
    }\hfill
    \subfloat[]{%
        \includegraphics[width=0.85\columnwidth]{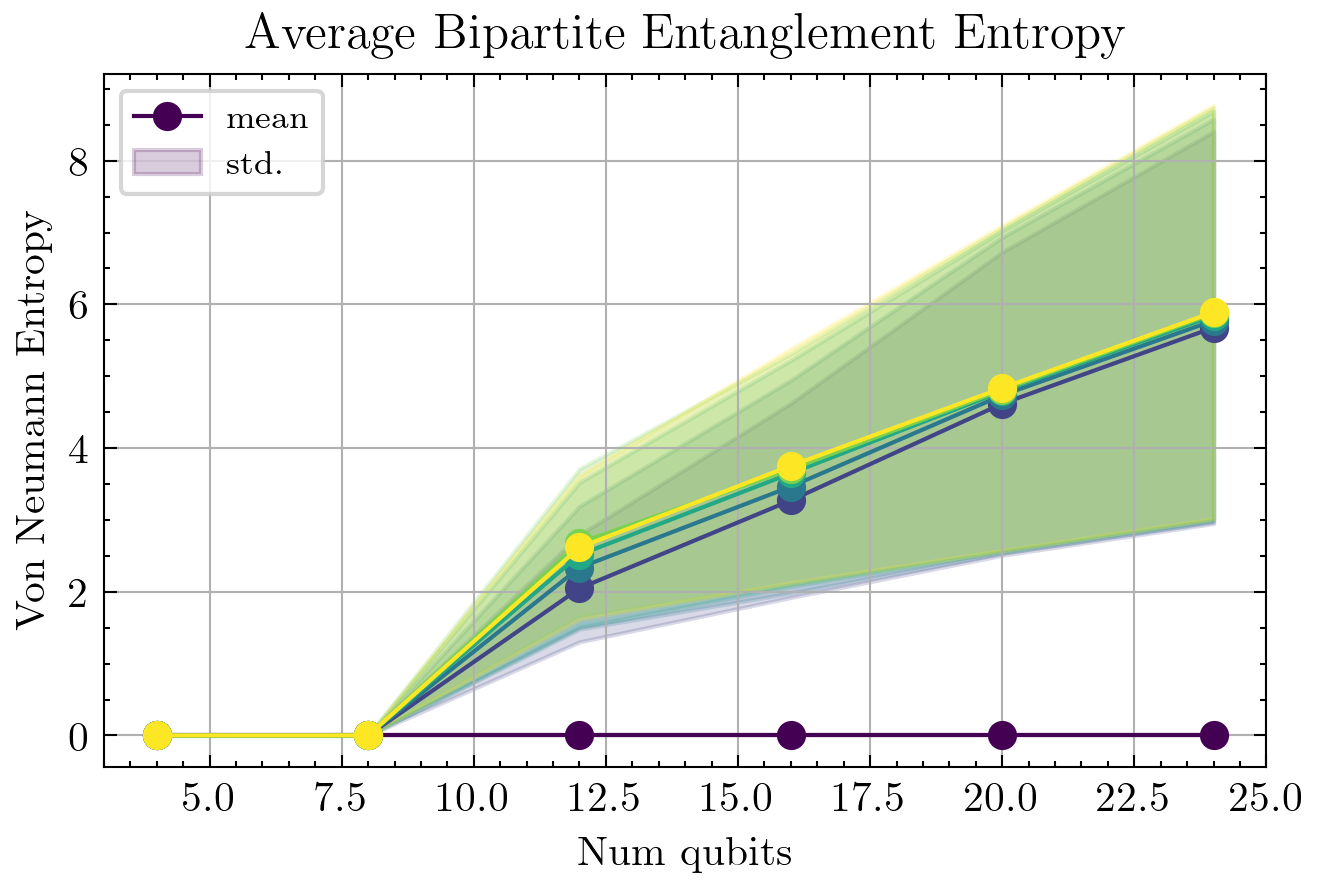}
    }
    \caption[]{The average (Panel (a)) and maximum (Panel (b)) Von Neumann entropy for bipartite splits along the direct qubit connections in MPS for a range of active space sizes (spin orbitals, hence the number of qubits). On Panel (b) the standard deviation (std.) was marked with semi-transparent shadow.}
    \label{fig:bip_ent_entropy}
\end{figure}

The results of maximum and average bipartite entropy over range of active space sizes in the Fig.~\ref{fig:bip_ent_entropy} shows the raise of the entanglement, with no signs of saturation of the entanglement entropy value. The entanglement in a system seems to rise linearly (except first two data points with almost zero entanglement). That could indicate exponential growth in bond dimension required to sustain accurate emulation over active space size applied in 1D tensor network representations, directly observed in the previous sections (\ref{fig:suff_bd_from_n_qubits}), clearly disobeying the area law for 1D systems. This observation is consistent with the expected limitation of one-dimensional MPS representations for time-evolved states generated by non-local molecular Hamiltonians. In the present benchmark, the measured entropy growth explains why the required bond dimension approaches the maximum value before reaching active-space sizes that are classically challenging for state-vector simulation. Although Fig.~\ref{fig:bip_ent_entropy_24q},~\ref{fig:bip_ent_entropy} reports entanglement entropy over multiple time steps, the full bond-dimension accuracy and runtime sweep was conducted at \(t=10\) a.u.; extending this sweep over full trajectories is left for future work.

\section{Discussion}
\label{sec:discussion}

\subsection{Scope and Limitations} 

The conclusions should be read as application-specific: the benchmark uses one DMET-embedded sulfonyl-fluoride fragment, Jordan--Wigner mapping, fixed orbital ordering, one first-order Trotterized circuit, and the observable evaluated at \(t=10\) a.u. The observed 20-24 qubit crossover is therefore not claimed to be universal across molecular systems, electronic structures, observables, mappings, orderings, or evolution times. Different molecules, active-space constructions, observables, and time horizons may exhibit different entanglement growth and therefore different MPS accuracy-runtime trade-offs. In particular, an entanglement-aware orbital ordering or an alternative fermion-to-qubit mapping could reduce the maximum entropy across some MPS cuts and shift the observed crossover to larger active spaces; testing such optimizations is outside the scope of the present benchmark. The accuracy criterion is task-specific to the one-body observable \(F(t)\), not global state fidelity or other observables, and the numerical behavior is reported for a single MPS emulator.

The study benchmarks MPS only; alternative geometries may change the accuracy-runtime trade-off and require separate evaluation.

The broader implication is therefore conditional rather than universal: when a molecular Hamiltonian-simulation workflow produces time-evolved states whose bipartite entanglement grows approximately linearly with active-space size in the chosen MPS ordering, the required bond dimension is expected to grow rapidly and may eliminate the practical advantage of MPS emulation. The present work demonstrates this behavior in one chemically motivated pharmaceutical benchmark and provides a protocol for testing whether the same limitation appears in related workflows.

\subsection{Implications for Pharmaceutical Hamiltonian Simulation Workflows}
\label{sub_sec:implications_for_drug_discovery}

The limitations observed in this study have implications for pharmaceutical Hamiltonian simulation workflows that rely on similar active-space construction, time evolution, and observable estimation:

1) Near-term quantum advantage: The challenges in classically emulating larger molecular systems underscore the potential value of even modest improvements in quantum hardware. Small increases in qubit count and coherence time could enable simulations beyond naive classical capabilities.

2) Hybrid approaches: Given the limitations of basic implementations of pure classical emulation, hybrid quantum-classical algorithms may offer a more promising near-term path to leveraging quantum effects in drug discovery workflows.

3) Algorithm design: results presented here were based on generic implementation of time evolution simulation, with no algorithmic improvements in regard to quantum circuit size. For a complete image of the usefulness of quantum emulators in this use case, the emphasis should be put on evaluation of quantum algorithms specifically designed to minimize circuit depth and simplify or reduce entanglement, potentially allowing more efficient classical emulation.

4) Resource allocation: Drug discovery organizations should carefully consider the trade-offs between investing in classical emulation techniques versus focusing on early quantum hardware access and algorithm development.

\subsection{Fundamental Limitations of Tensor Network Representations} 

The observed scaling behavior displays a practical limitation of one-dimensional MPS emulation for the molecular time-evolution states generated in this benchmark. Rapid growth in the required bond dimension appears to be intrinsically linked to the entanglement structure of molecular systems, particularly those that involve delocalized electronic states common in pharmaceutical compounds. This known aspect of the classical simulation of quantum systems is confirmed by the results of this work. The runtime scaling for accurate target-observable estimation with MPS emulation in this use case reflects a fundamental challenge in representing highly entangled quantum states using classical tensor network structures.

The trade-off between accuracy and computational efficiency becomes particularly acute when considering characteristics of quantum chemistry applications:

\begin{enumerate}
    \item The non-local nature of electronic correlations in molecular systems
    \item The complex orbital interactions involved in chemical reactivity
    \item The need for high precision in energy (or other properties of the system) calculations for meaningful chemical predictions

\end{enumerate}

On the other side, tensor networks are powerful tools for accurate simulation of quantum systems displaying low or simply structured entanglement, e.g. with local interactions only obeying area laws,  or with multiscale entanglement structures, where entanglement is system scale-invariant. Molecular Hamiltonian-simulation workloads do not, in general, guarantee the low-entanglement structure required for efficient one-dimensional MPS compression, especially after real-time evolution under non-local fermionic interactions.

\section{Conclusion}
\label{sec:conclusion}

We benchmarked MPS tensor-network emulation of a one-step first-order real-time molecular Hamiltonian-evolution circuit for a DMET-embedded pharmaceutical fragment across active spaces of 4-24 qubits, measuring runtime, accuracy, and entanglement growth under a chemical-accuracy--derived precision criterion.

The results show that the bond dimension required for accurate estimation of the chosen temporal observable grows rapidly with active-space size, reaching the maximum MPS bond dimension at the largest systems tested. Although MPS emulation scales favorably at fixed bond dimension, this advantage is lost once the bond dimension needed for accuracy grows with the entanglement of the time-evolved state. The entropy analysis supports the expected interpretation: the one-dimensional MPS structure is poorly matched to the non-local entanglement generated by this molecular Hamiltonian workflow. 

For this fixed-ordering, observable-level benchmark, MPS emulation is therefore useful only in the smaller active spaces, which are also accessible to state-vector simulation, and provides no practical advantage in the 20-24 qubit regime once accuracy constraints are enforced. The observed crossover should not be interpreted as universal, but as a concrete engineering boundary for this molecule and for workflows that exhibit similar entanglement growth. Extending this boundary may require entanglement-aware algorithm design, alternative tensor-network geometries, hybrid workflows, or fault-tolerant quantum hardware; these alternatives were not benchmarked here.

\section{Implications and Design Directions}
\label{sec:implications_design_directions}

The results presented in this work clarify not only the limitations of MPS–based tensor network emulation for molecular Hamiltonian simulation, but also the design space for future emulator and algorithm development. In particular, the observed entanglement-driven growth in bond dimension provides concrete guidance on which directions are most likely to yield practical improvements and which are fundamentally constrained.

\subsection{Alternative Tensor Network Geometries}

The breakdown of MPS efficiency at chemically relevant active space sizes highlights the importance of tensor network geometries that better match the entanglement structure of molecular electronic states. Tree tensor networks (TTN)\cite{vidal2008class} and projected entangled pair states (PEPS)\cite{verstraete2004renormalization} are natural candidates, as they relax the one-dimensional constraint inherent to MPS and may more efficiently capture multi-dimensional and non-local correlations. 

Recent algorithmic advances also show promise for improving tensor network performance. \cite{alkabetz2021tensor} demonstrated that belief propagation algorithms can accelerate tensor network contraction, while \cite{evenbly2023loop} developed loop series expansions that could improve the efficiency of representing highly entangled states. \cite{gray2021hyper} showed that hyper-optimized tensor network contraction can significantly reduce computational costs, and \cite{pan2024efficient} achieved substantial performance improvements through GPU acceleration of tensor operations.

However, those potential benefits must be weighed against increased algorithmic and contraction complexity. From a systems perspective, the key question is not whether such geometries are more expressive in principle, but whether they can deliver improved accuracy–runtime trade-offs for realistic molecular workloads within practical resource limits. Direct comparison against these architectures is left for future work.

\subsection{Hybrid Classical--Quantum Workflows}
The engineering boundary identified in this study suggests a complementary role for hybrid classical--quantum approaches. In such workflows, tensor network methods could be applied selectively to weakly entangled subsystems or as preprocessing tools, while strongly entangled components of the simulation are delegated to quantum hardware. This division aligns naturally with embedding-based quantum chemistry methods and offers a pragmatic pathway to extend simulation scales prior to the availability of fully fault-tolerant quantum computers. Importantly, the results here help delineate where such partitioning becomes necessary rather than optional.

\subsection{Algorithm and Emulator Co-Design Considerations}
Finally, the findings underscore the importance of co-design between quantum algorithms and emulation strategies. The benchmarks presented here relied on generic implementations of Hamiltonian time evolution, without explicit optimization to suppress entanglement growth or reduce circuit depth. For emulator-assisted workflows to remain viable, future algorithmic designs must explicitly account for their entanglement footprint, contraction cost, orbital ordering, and fermion-to-qubit mapping. From an engineering standpoint, this shifts the emphasis from standalone emulator performance to integrated algorithm–emulator system design.

All the directions listed above follow directly from the measured scaling behavior reported in this work and outline realistic paths for extending quantum simulation capabilities. While no single classical approach is likely to overcome the fundamental entanglement barriers identified here, informed system-level design choices can help bridge the gap between near-term emulation and long-term quantum advantage for molecular simulation.

\section{Authors contribution}

M.K. led the technical execution of the study, including the design and implementation of the emulation experiments and the primary data analysis. P.L. provided overall scientific direction and contributed to the interpretation of results and manuscript writing. E.M. contributed to implementation, validation, and analysis in the early stages of the work. P.P. contributed to conceptual framing and interpretation of results. All authors reviewed and approved the final manuscript.

\section{Acknowledgments}

The authors thank Eline Welling and Ido Niesen (Fermioniq) for providing computational resources, for helpful discussions that shaped this study, and for feedback on our use of the Fermioniq Ava emulator. We also thank Matthew Pigg and Capgemini Engineering's Cortex programme of projects developing capabilities in new and emerging technologies, within which the initial research was conducted. Finally, we acknowledge Ivan Rungger and Lachlan Lindoy for stimulating discussions and idea exchange.

\section*{Appendix A. Expectation values of measurement operator}

\begin{figure}
    \centering
    \subfloat[]{%
        \includegraphics[width=0.95\columnwidth]{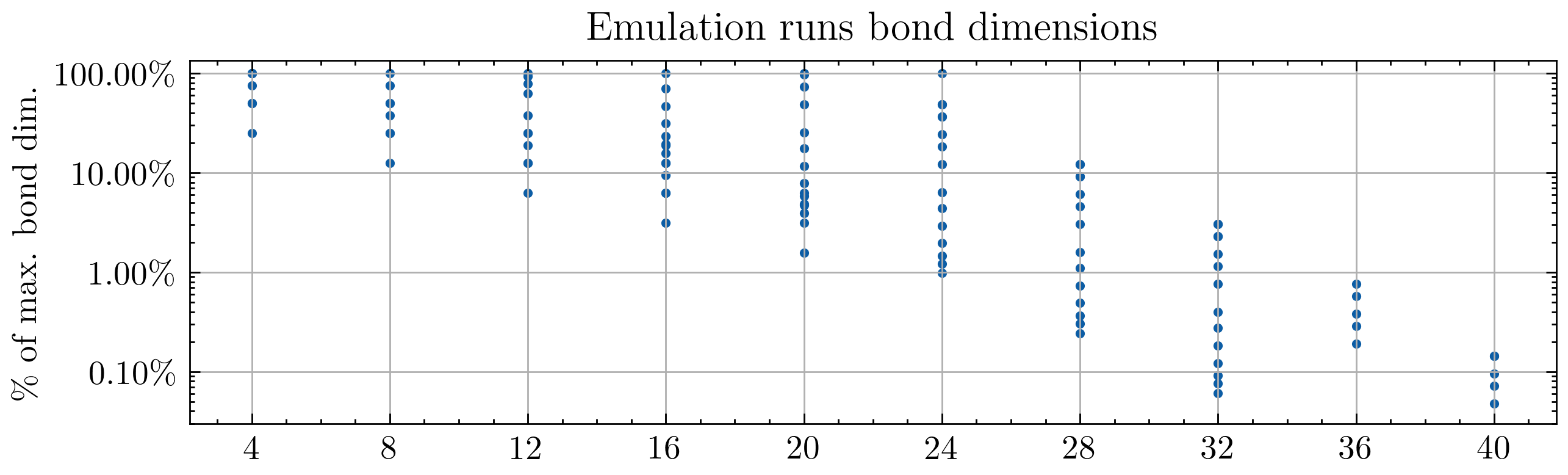}
    }\hfill
    \subfloat[]{%
        \includegraphics[width=0.95\columnwidth]{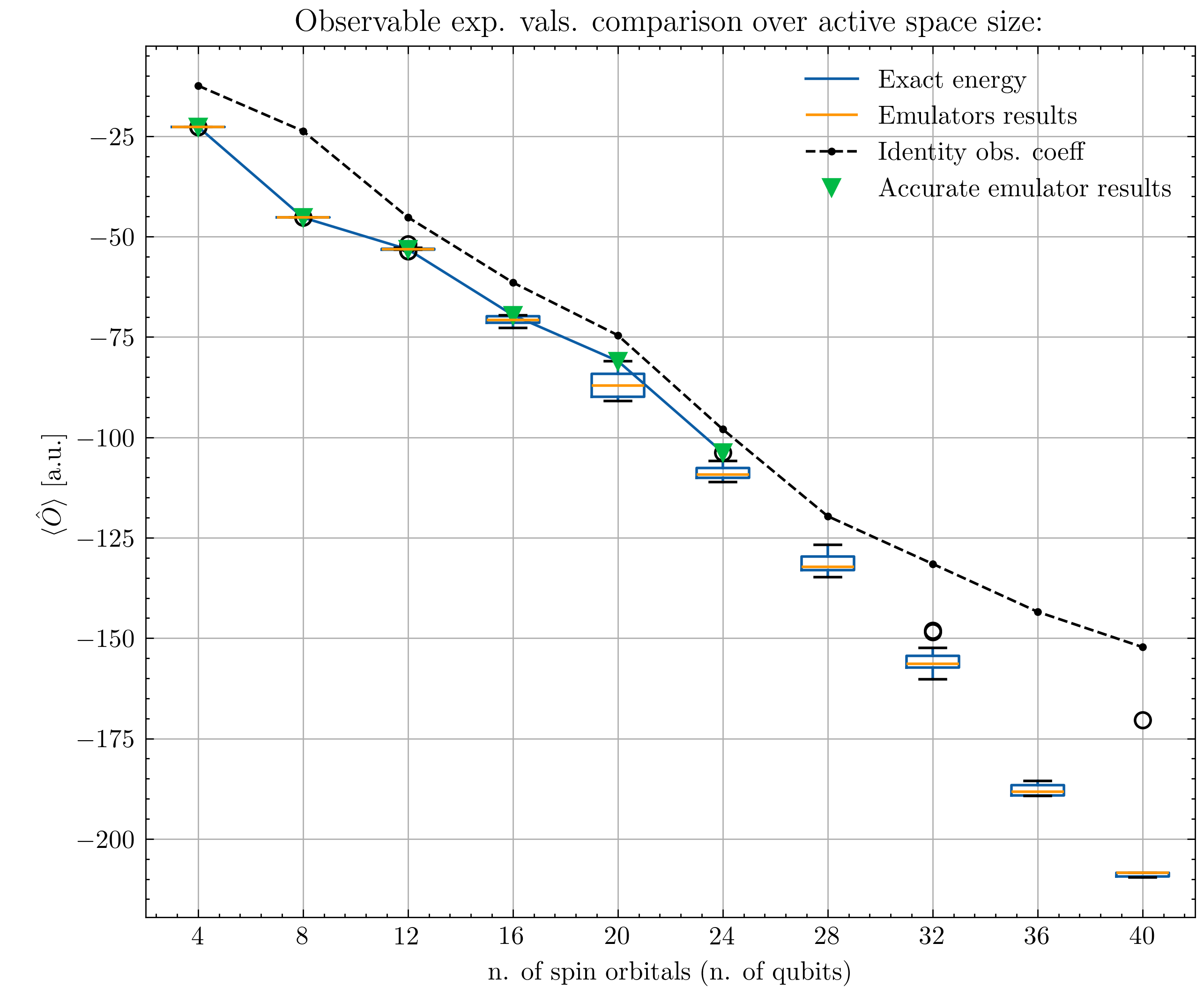}
    }
    \caption[]{Observable expectation values spread from time-evolution circuits emulation for $t=10$ \textit{au} over active space sizes (panel (b)). The spread was showcased with a boxplot, where the orange line is the median, blue box shows the range between the first and third quartile. Error bars extend this range with the farthest data point lying within 1.5 times the inter-quartile range (IQR) from the box. The rest of the data points are plotted with empty black circles. Additionally, the accurate results from the state-vector simulator was added up to 24-qubits with the green triangle. As a point of reference, identity observable coefficient was plotted. The bond dimension for which the emulations were run are plotted in panel (a).}
    \label{fig:time_evolution_energies_spread}
\end{figure}

The spread of the temporal observable expectation values that we can expect from the MPS emulators for a range of bond dimensions is plotted in Fig.~\ref{fig:time_evolution_energies_spread}. The spread varies from the number of qubits, but it is always definitely below the identity term coefficient level and interestingly it is also below or at the similar level as the accurate emulators results. 

We do not use global state fidelity as a selection criterion in this benchmark. Preliminary fidelity diagnostics were found to be sensitive to emulator state export and normalization details, and are therefore not reported. The accuracy claims in the main text are restricted to the target observable \(F(t)\) compared against state-vector ground truth.

\noindent

\section*{Appendix B. Emulation runs details}
\label{app_sec:emulation_run_details}

\begin{figure}
    \centering
    \includegraphics[width=0.75\linewidth]{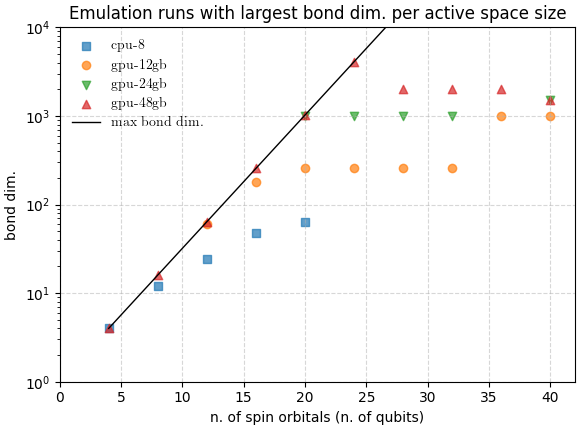}
    \caption{Emulation runs with the largest bond dimension per active space size for different types of backends.}
    \label{fig:bd_runs_per_backend}
\end{figure}

During this study, we performed emulations across a range of active-space sizes and classical backend types.
Capacity and efficiency of the backend required to run the emulation of given circuit and bond dimension, may be hard to assess. To aid reproducibility, in Fig.~\ref{fig:bd_runs_per_backend} are plotted (per each backend type) the maximum bond dimensions that the experiments were run during this study. Each of this data points denotes successful, finished emulation. It should be noted that, for given types of backend and active space size, the larger bond dimension emulations might be possible to run, but they were not, because of impractical estimated runtime.

\noindent
\FloatBarrier


\bibliographystyle{IEEEtran}
\bibliography{IEEEabrv,bib} 

\end{document}